%% file: main.tex
\pdfoutput=1
\documentclass[11pt]{article}

\usepackage[final]{coling}

\usepackage{times}
\usepackage{latexsym}

\usepackage[T1]{fontenc}

\usepackage[utf8]{inputenc}

\usepackage{microtype}

\usepackage{inconsolata}

\usepackage{graphicx}
\usepackage{pifont}
\usepackage{natbib}
\usepackage{booktabs,multirow}
\usepackage{float}
\usepackage{enumitem}

\newcommand{\cmark}{\ding{51}}%
\newcommand{\xmark}{\ding{55}}%

\newcommand{\dataset}{\texttt{ACL-rlg}}

\title{\dataset{}: A Dataset for Reading List Generation}

\author{
  \textbf{Julien Aubert-Béduchaud\textsuperscript{1}},
  \textbf{Florian Boudin\textsuperscript{1,2}},
  \textbf{Béatrice Daille\textsuperscript{1}},
  \textbf{Richard Dufour\textsuperscript{1}}
  \\
  \textsuperscript{1} Nantes Université, École Centrale Nantes, CNRS, LS2N, UMR 6004, F-44000 Nantes, France
  \\
  \textsuperscript{2} JFLI, Tokyo, Japan
  \\
  \small{
    \textbf{Correspondence:} \href{mailto:julien.aubert-beduchaud@univ-nantes.fr}{julien.aubert-beduchaud@univ-nantes.fr}  
    }
}

\begin{document}
\maketitle
\begin{abstract}

Familiarizing oneself with a new scientific field and its existing literature can be daunting due to the large amount of available articles. Curated lists of academic references, or \textit{reading lists}, compiled by experts, offer a structured way to gain a comprehensive overview of a domain or a specific scientific challenge. In this work, we introduce \dataset{}, the largest open expert-annotated reading list dataset. We also provide multiple baselines for evaluating reading list generation and formally define it as a retrieval task. Our qualitative study highlights the fact that traditional scholarly search engines and indexing methods perform poorly on this task, and GPT-4o, despite showing better results, exhibits signs of potential data contamination.

\end{abstract}

\section{Introduction}
\label{sec:introduction}

%
As the volume of scientific publications continues to grow, gaining insights into a field becomes increasingly time-consuming. 
Although existing tools for browsing the literature (academic search engines, paper recommendation systems, etc.) help researchers avoid missing relevant papers, they often return an overwhelming number of results.
This abundance of information makes the familiarization process daunting and inefficient, particularly for junior researchers who lack effective paper-skimming skills or experienced scholars transitioning to a new field.
%

%
%

One solution is to consult survey papers, which offer comprehensive reviews of the current state of research in a particular area.
However, survey papers have several limitations: they are not available for all fields, may become outdated quickly, and may not address the specific needs of novice researchers as surveys are too broad and without explicit instructions of reading order.
Another way to familiarize oneself with a new field is through \textit{reading lists}, which are curated lists of academic references compiled by experts that provide an organized overview of a field~\citep{Siddall_2014}.
Compared to surveys, creating reading lists requires significantly less effort, yet they can still help novice researchers navigate key literature and reduce the time needed to begin their research~\citep{UCAM-CL-TR-848}.
As a result, interest in automated methods for generating reading lists has increased~\citep{10.1145/1864708.1864740, UCAM-CL-TR-848, gordon-etal-2017-structured, 10.1109/JCDL.2019.00011}.
%
%
However, progress has been limited due to the scarcity and quality of available datasets.
%
%
Indeed, existing datasets are either too small, as exemplified by the eight lists compiled by~\citet{UCAM-CL-TR-848}, or are of low quality because they have been automatically constructed from survey papers~\citep{10.1145/1864708.1864740,9835398,10.1109/JCDL.2019.00011}.

%
%
To address this gap, we introduce \dataset{}, a new dataset of expert-curated reading lists derived from tutorials accepted at major Natural Language Processing (NLP)-related conferences, along with manually annotated queries that novice researchers might use to locate relevant papers from the lists.

%
We conducted experiments with various methods for generating reading lists, including academic search engines, commercial LLMs, and ad-hoc retrieval models, to validate our dataset and establish a benchmark for future research.
Our contributions are:
%
\begin{itemize}[itemsep=0.3em,topsep=0.3em]
    \item We introduce \dataset{}, the largest expert-curated reading list dataset available with 85 reading lists;
    \item We formally define the reading list generation task as a retrieval task and set up an evaluation framework with metrics and baselines;
    \item We compare existing models and show that there are signs of potential data contamination using systems such as GPT-4o.
\end{itemize}

Our code and data are openly available\footnote{ \href{https://github.com/jjbes/aclanthology-tutorial-reading-lists}{github.com/jjbes/aclanthology-tutorial-reading-lists}}.

\section{Reading List Generation}
\label{sec:reading-list-generation}

In the literature, \emph{reading list generation} is broadly defined as the task of creating a list of references that serves as a starting point for familiarizing oneself with a new field~\citep{10.1145/1864708.1864740, UCAM-CL-TR-848, gordon-etal-2017-structured, sesagiri_raamkumar_using_2017, 10.1109/JCDL.2019.00011}.
The lack of a standardized definition for the task makes it challenging to compare the results between these studies, as the proposed models operate with varying inputs and outputs.
%
Here, we address this issue by defining reading list generation as an article retrieval task:
\begin{itemize}[topsep=.3em,leftmargin=.8em]
\item[] Given a collection of scientific articles $C=\{a_1, a_2, \cdots, a_N\}$ and a query $q$ formulated by a novice researcher (e.g., using keywords or a natural language expression), the task of reading list generation is to retrieve a concise, ordered list of papers $L$ that helps the user efficiently grasp the topic of $q$. The list should be compact for quick consumption, with an order that ideally supports the user’s learning curve.

\end{itemize}

Reading lists should offer an overview of a research field while remaining concise, balancing the relevance of references with general coverage of the existing literature~\citep{2436/3693}.
Although the maximum size of a reading list is commonly set at 20 articles~\citep{UCAM-CL-TR-848, sesagiri_raamkumar_using_2017, 10.1109/JCDL.2019.00011, 9835398}, there is no consensus on the minimum number of references.
Based on the requirements outlined by~\citet{sesagiri_raamkumar_using_2017}, we assume that a reading list should include at least three references.
%


One notable aspect that distinguishes the generation of reading lists from other ad hoc retrieval tasks is the meaningful ordering of the papers within the lists.
In a reading list, articles are arranged in a sequence that reflects the optimal learning path, guiding the reader through a knowledge progression to better understand the new field~\cite{9835398}.
Therefore, generating a reading list not only involves retrieving relevant papers but also determining the best order in which to present them.


%

Benchmark datasets for reading list generation are scarce.
Most existing datasets are constructed using references from survey papers as proxies for reading lists.
For instance, \citet{10.1145/1864708.1864740} used survey papers from the ACM Computing Surveys, \citet{10.1109/JCDL.2019.00011} relied on reviews and survey papers from Scopus, and \citet{9835398} collected survey papers by querying S2ORC~\cite{lo-wang-2020-s2orc} and Google Scholar.
It is worth noting the lack of uniformity in the queries associated with these reading lists.
\citet{10.1145/1864708.1864740} used a set of initial papers as a query, while \citet{10.1109/JCDL.2019.00011} and \citet{9835398} relied on keywords, either provided by the survey authors or extracted from their titles, respectively.
These automatically constructed datasets can be extended to a large scale, with~\citet{9835398} aggregating more than 9,000 reading lists.
However, their quality is limited as the purpose of surveys is to be comprehensive, which contrasts with the conciseness and focus required in reading lists.

Perhaps the most relevant work for us is by~\citet{UCAM-CL-TR-848}, who introduced a dataset of reading lists created by experts using the ACL Anthology Network~\citep{radev_acl_2013}.
NLP-related PhDs with several years of research experience were tasked with creating reading lists from research topics (e.g.~\textit{statistical parsing}).
While the resulting dataset is high in quality, its small size (only 8 lists) and its age significantly render it less useful for contemporary research, showing the need for newer, more robust datasets for reading list generation. 

\begin{table*}[htbp]
    \centering
    \resizebox{.9\linewidth}{!}{
        \input{tables/datasets-compact}
    }
    \caption{Comparison of existing reading list evaluation datasets.}
    \label{tab:datasets}
\end{table*}

\section{Dataset Description}
\label{sec:dataset-description}

\subsection{Reading Lists Extraction}
\label{subsec:data-sourcing}

We gathered reading lists from tutorials presented at events sponsored by the Association for Computational Linguistics (ACL).
Tutorials are structured educational sessions presented by experts and designed to provide in-depth guidance in a specific field.
Since 2021, tutorial presenters have been explicitly instructed to include a reading list in their description papers\footnote{\href{https://2021.aclweb.org/calls/tutorials/}{https://2021.aclweb.org/calls/tutorials/}}.
These instructions suggest that the lists should be concise, recommending 4-10 articles, and include a range of authorship to ensure diverse perspectives.
We reviewed all tutorial descriptions and identified those containing sections explicitly titled ``Reading List'' or ``Prerequisite Readings''. 
References were manually extracted from these sections, and their metadata was enriched using the Semantic Scholar API. 
We maintained the original ordering of the references and preserved any structural organization provided, such as sections or subsections. 
A total of 27 reading lists included structured sections, and among those, 3 included subsections (see Table~\ref{tab:section_count} in Appendix). 

\begin{figure}[htbp]
    \centering
    \includegraphics[width=\linewidth]{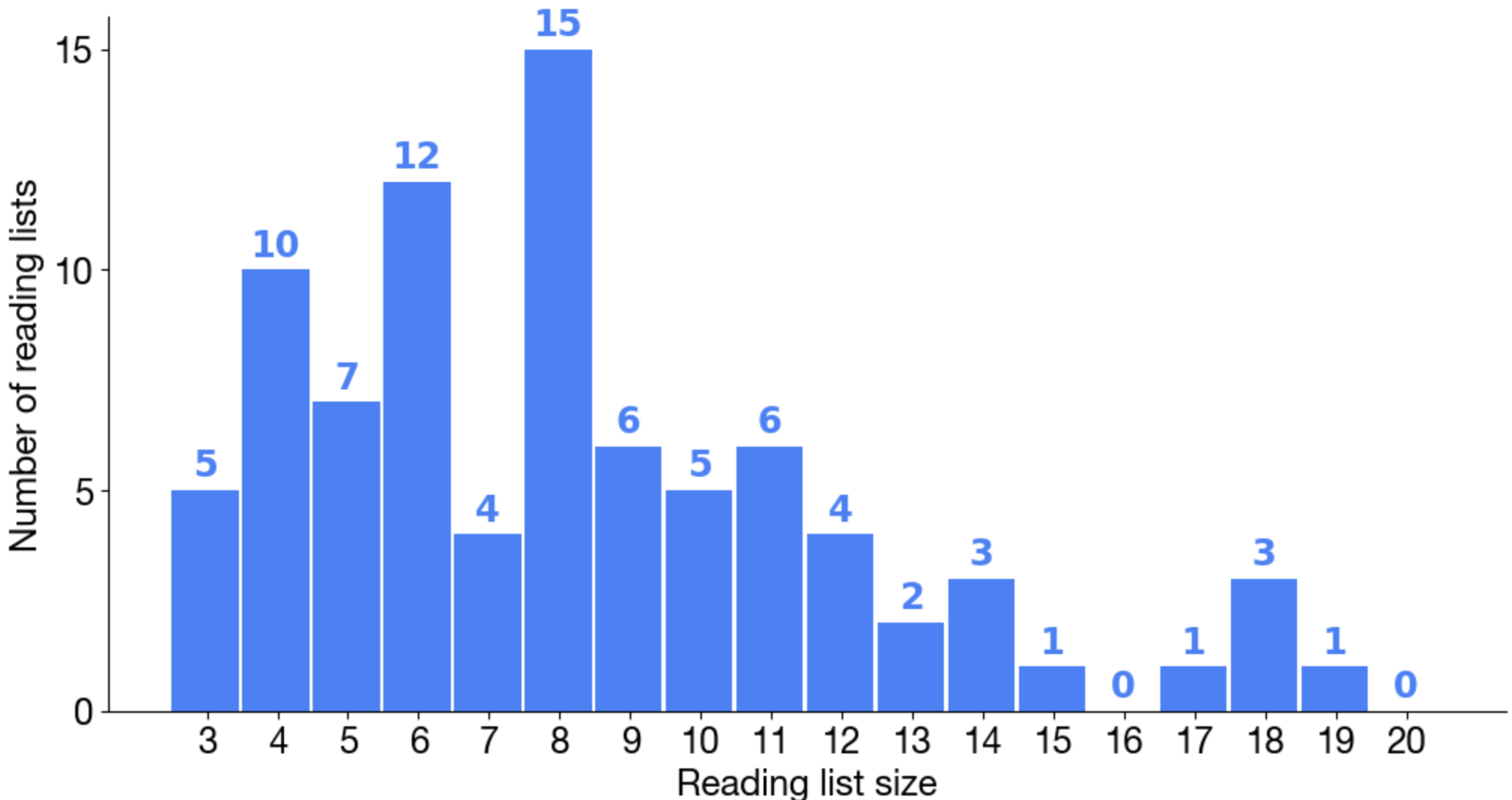}
    \caption{Distribution of tutorials by their number of reading list references.}
    \label{fig:list-length}
\end{figure}

Although the structural organization was not used in current experiments, it may serve as a valuable resource for future models aiming to capture specific aspects of reading lists (e.g. Methods, Datasets, etc.). However, because the ACL instructions do not specify any particular rules for ordering, we cannot assume these lists are optimally arranged.
Finally, we filtered the collected reading lists by length, keeping reading lists within 3 to 20 papers (see Figure \ref{fig:list-length} for detailed statistics). 

The resulting dataset consists of \textbf{85 expert-crafted reading lists}, citing a total of 662 distinct papers.
The publication dates of the referenced works range from 1977 to 2023, with the majority of papers published between 2016 and 2023, as shown in Figure~\ref{fig:histogram_year}.
Table~\ref{tab:datasets} provides a comparison between \dataset{} and existing datasets for reading list generation.

\subsection{Query Annotation}
\label{subsec:query-annotation}

\noindent \textbf{Annotation protocol.}
In the context of automatic reading list generation, an initial query is required to retrieve a set of relevant references.
To achieve this, we need to formulate a query that accurately reflects the field of the tutorial from which the reading list was derived.
A simple approach might involve extracting keywords directly from the tutorial titles.
However, titles may omit critical concepts since they are crafted primarily to capture attention. 
Instead, we opted for a more controlled and qualitative approach, asking annotators to create queries they would use to find a reading list relevant to the field of each tutorial, with novice researchers as target users (See annotation guidelines in Appendix~\ref{appendix:annotation}).

\begin{table}[htbp]
    \centering
    \resizebox{\linewidth}{!}{%
    \input{tables/query-example}
    }
    \caption{Queries for the~\citet{glavas-etal-2019-computational} tutorial.}
    \label{tab:example}
    \vspace{-.5em}
\end{table}

For each reading list, three annotators --each with varying levels of NLP expertise (a PhD student (A1), a postdoc (A2), and a senior researcher (A3))-- were asked to create keyword-based queries.
Additionally, we tasked them with providing natural language instructions for an LLM-based assistant.
An example of queries is presented in Table~\ref{tab:example}, with notable differences observed between annotators.

\begin{table}[htbp]
    \centering
    \resizebox{\linewidth}{!}{%
    \input{tables/exp1-keywords}
    }
    \caption{Inter-annotator agreement on keyword queries.}
    \label{tab:exp1-keywords}
    \vspace{-.5em}
\end{table}

\noindent \textbf{Inter-annotator agreement.}
Table~\ref{tab:exp1-keywords} shows the inter-annotator agreement, measured by the exact word match between queries using stemming and lowercase. 
A key observation is the significant difference in query length, particularly between A2 and the other two annotators, which partly explains the mismatch.  
The discrepancies may also arise from the fact that some terms annotated in the queries do not appear in the tutorial descriptions, introducing variability when annotating similar topics. 
This trend is also apparent in the natural language instructions (see Table~\ref{tab:exp1-sentences} in Appendix). 

We also compared the manually generated queries with those produced by FirstPhrases, an automatic keyword extraction method that extracts noun phrases from tutorial titles~\cite{boudin:2016:COLINGDEMO}.
We set the number of extracted keywords to three, following \citet{LI2017666} average query length commonly used in academic search.
We notice a moderate word overlap between automated and manual queries ($\approx$40\%), highlighting the inappropriateness of titles as a proxy for queries.

\section{Experiments}

%

\subsection{Results of Search Engines and LLMs}
\label{search-engines-comparison}

\noindent \textbf{Methodology.}
As researchers often rely on academic search engines to find relevant articles, we investigate whether the top results from these engines align with expert-curated reading lists.
%
Specifically, we compare the outputs of two widely used search engines: Semantic Scholar (S2) and Google Scholar (GS). 
%
Given the growing popularity of LLMs as an alternative, we also measure GPT-4o~\cite{openai2024gpt4technicalreport} and Gemini 1.5~\cite{geminiteam2024gemini15unlockingmultimodal} performances against these search engines for the same task with different output strategies (Basic, JSON-mode (JM), Structured Outputs (SO)). The temperature of LLMs was set to 0 in our experiments.
%
We evaluate the generated reading lists using commonly-used IR metrics: Recall@k, NDCG@k (Normalized Discounted Cumulative Gain), and MRR@k (Mean Reciprocal Rank), with k set to 20. Recall measures the model's ability to predict reading lists that closely match those curated by experts. In contrast, NDCG and MRR assess the quality of the ranking within the predicted lists, focusing on how well the relevant items are ordered.
Search engines are queried using manually annotated keywords, while LLMs are prompted with instruction-based queries, following the procedures detailed in Appendix~\ref{appendix:requests}.\\

\noindent \textbf{Analysis of results.}
As shown in Table~\ref{tab:current-methods-short} (full details in Table~\ref{tab:current-methods}), GPT-4o outperforms other systems in generating reading lists.
Notably, GPT-4o demonstrates significant improvements in prioritizing articles from expert-curated lists, as reflected by its MRR@20 scores.
However, the overall scores remain low, highlighting the need for systems dedicated to this task. 
A closer examination of individual annotator scores reveals that the queries annotated by the senior researcher yield the highest scores. Despite their shorter length, these queries seem to be the most carefully constructed. 
Regarding the lower performances of models on A2 annotations, we hypothesize that A2's longer, more precise queries may overly narrow the search scope, thus negatively impacting retrieval effectiveness. 
An in-depth analysis of the differences between queries formulated at various levels of expertise could provide further insights into this issue, which we leave for future work.

\begin{table}[htbp]
    \centering
    \resizebox{\linewidth}{!}{%
        \input{tables/current-methods-short}
    }
    \caption{Performance of Search Engines and LLMs on the reading list generation task.}
    \label{tab:current-methods-short}
\end{table}

\subsection{Comparison of Retrieval Models on ACL Anthology Collection}

\noindent \textbf{Additional experiments.}
We conducted additional experiments by indexing the collection of papers from the ACL Anthology. 
We compared three retrieval systems: Semantic Scholar (Any article from the S2 collection for comparison purpose, ACL articles only, and most cited ACL articles) as a baseline, as well as sparse (BM25~\cite{10.1561/1500000019}) and dense (SPECTER2~\citep{singh-etal-2023-scirepeval}) retrieval models (see Table~\ref{tab:classic-methods-short}, full details in Table~\ref{tab:classic-methods}).
Overall, we observe lower scores compared to those achieved with search engines but found performance comparable to Semantic Scholar when using the same index.\\

\begin{table}[htbp]
    \centering
    \resizebox{\linewidth}{!}{%
        \input{tables/classic-methods-short}
    }
    \caption{Performance of Retrieval Models on the reading list generation task.}
    \label{tab:classic-methods-short}
\end{table}

\begin{figure}[ht!]
    \centering
    \includegraphics[width=1\linewidth]{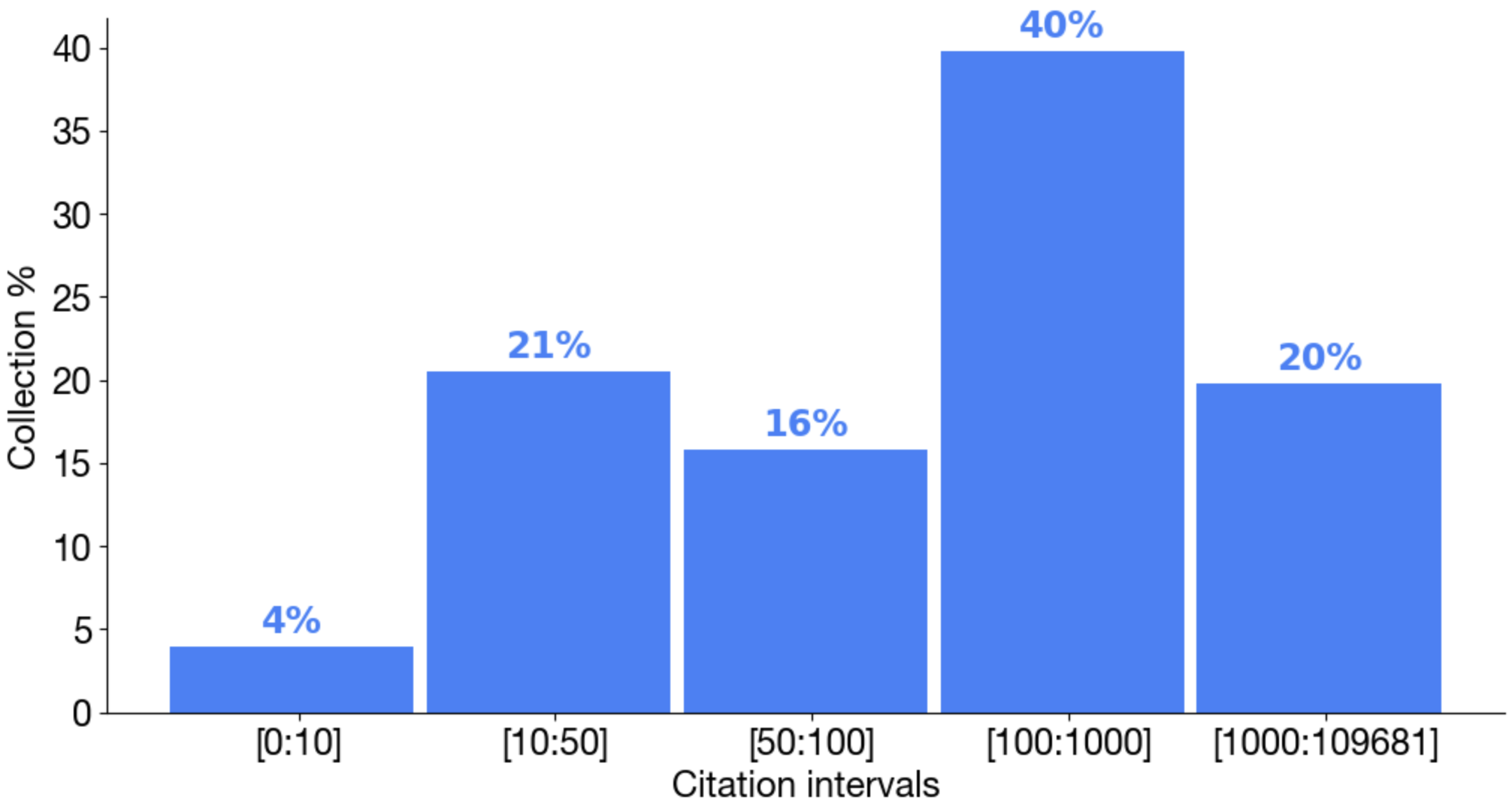}
    \includegraphics[width=1\linewidth]{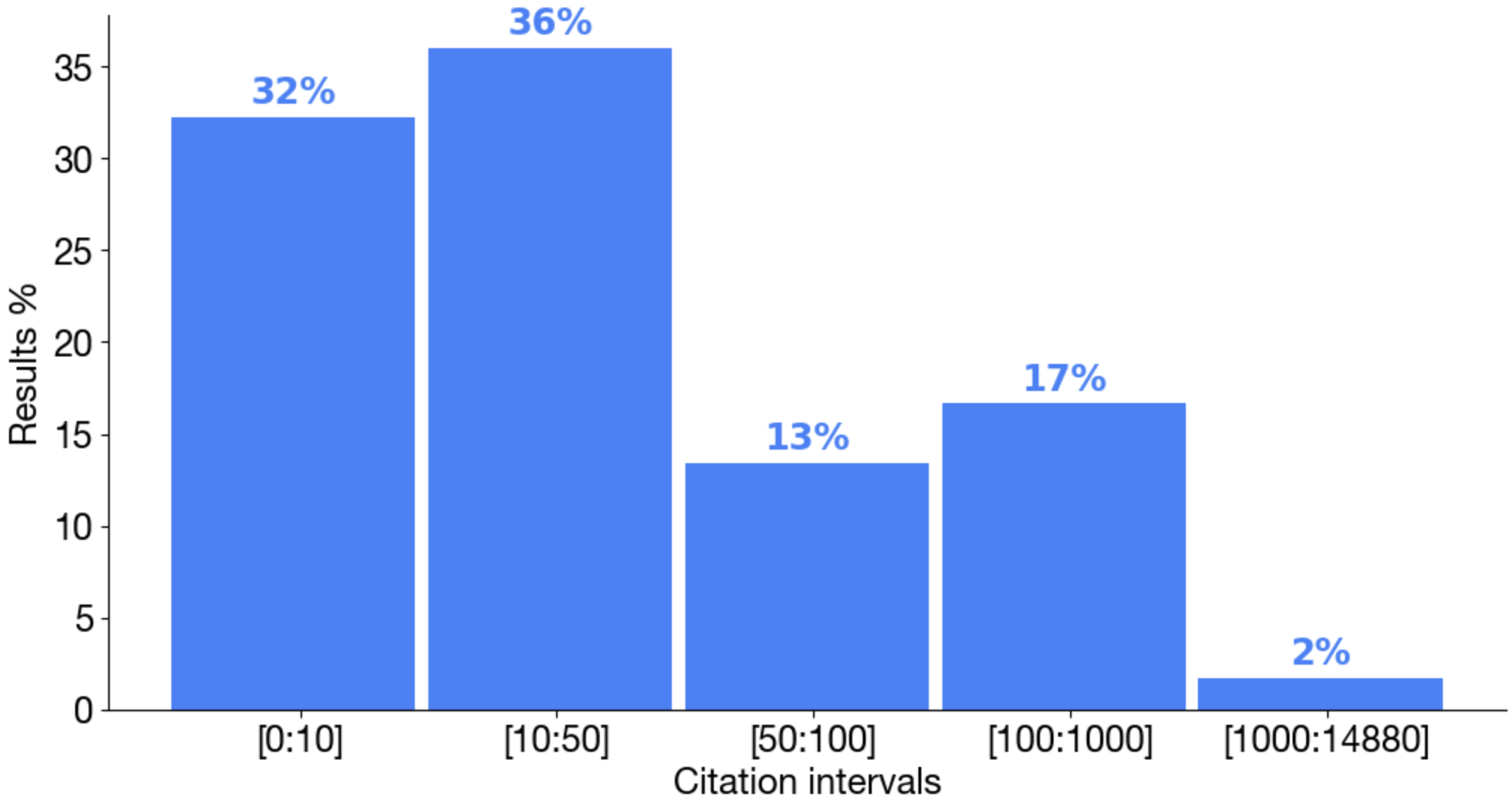}
    \caption{Comparison of overall article popularity trends within the dataset reading lists (TOP) and the corresponding predicted outcomes (BOTTOM).}
    \label{fig:citations}
\end{figure}

\noindent \textbf{Article popularity impact.}
Interestingly, the predicted results from Semantic Scholar exhibit popularity trends that are significantly lower compared to the articles featured in the \dataset{} reading lists (Figure~\ref{fig:citations}).
Given these differences, we derived additional results from Semantic Scholar using a "Most Cited" strategy, selecting the 20 most-cited articles from the 100 relevant results for a given query.
The performance improvement observed with this strategy suggests that prioritizing highly cited articles could be an important factor to consider when creating a reading list for a specific field.

\subsection{Analysis of LLMs Predictions}

\begin{figure}[htbp]
    \centering
    \includegraphics[width=1\linewidth]{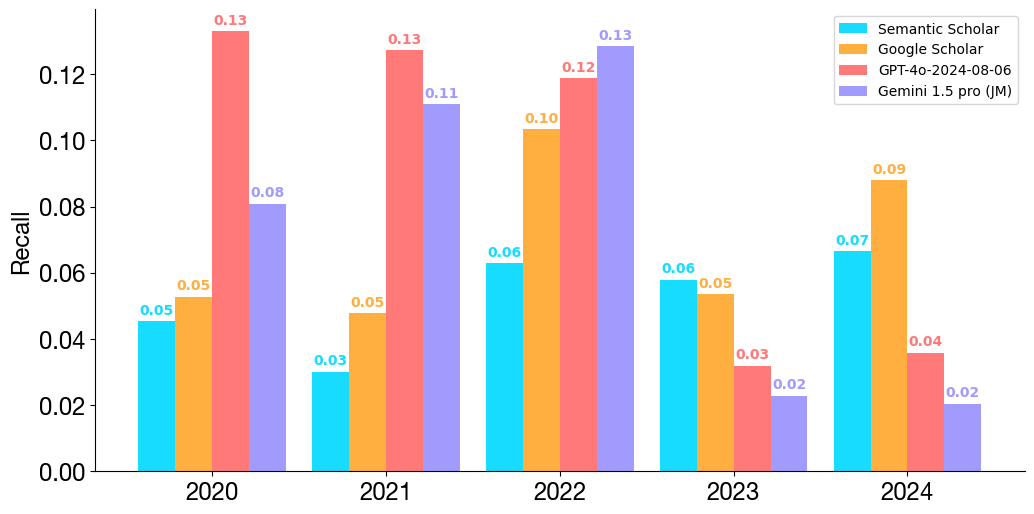}
    \caption{Recall@20 scores for each year of tutorials from which the reading lists were extracted.}
    \label{fig:gpt4o_recall}
    \vspace{-.5em}
\end{figure}

\textbf{Data contamination.}
Given that some of the tutorials used to extract our reading lists may be part of GPT-4o's training data (up to October 2023), the higher scores achieved by GPT-4o raise concerns about potential data contamination with LLMs~\cite{sainz-etal-2023-nlp}.
To gain insights into this issue, we evaluated the performance of search engines and LLMs based on the reading lists extracted from tutorials for each specific year (see Figure~\ref{fig:gpt4o_recall}).
While there is no definitive evidence, our results show a drop in performance for LLMs from 2023 that is not observed with the search engines, strongly suggesting data contamination.  \\

\begin{figure}[htbp]
    \centering
    \includegraphics[width=1\linewidth]{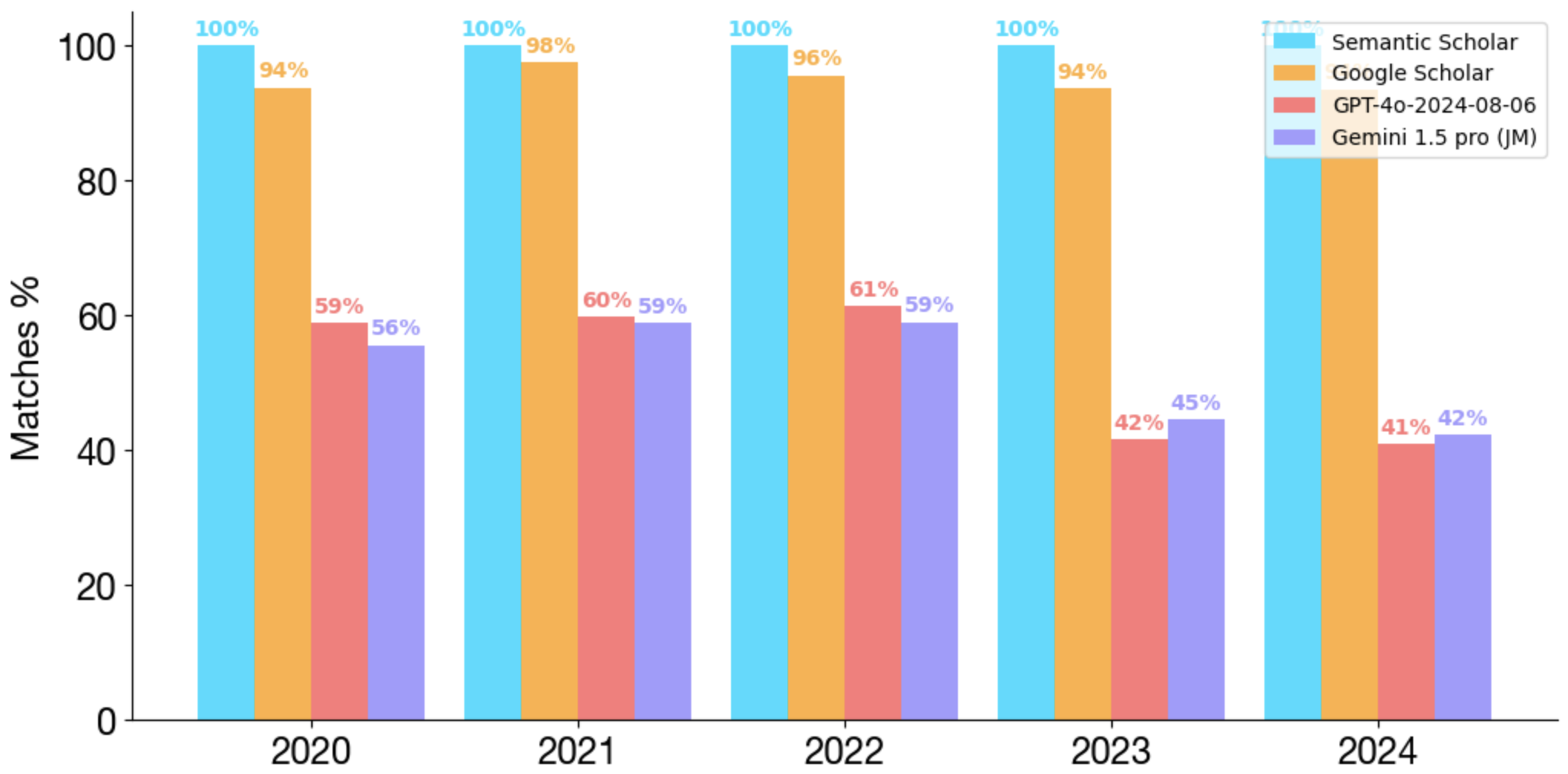}
    \caption{\% of references in Semantic Scholar.}
    \label{fig:gpt4o_match}
    \vspace{-.5em}
\end{figure}

\noindent \textbf{Hallucinations.}
Another concern with the use of LLMs is the potential generation of hallucinated references.
Here, we consider references that are not indexed in Semantic Scholar as hallucinations.
Figure~\ref{fig:gpt4o_match} shows the percentage of hallucinated references in the generated reading lists for each year.
We observe that $\approx\frac{1}{3}$ of the references generated by LLMs could not be found in Semantic Scholar, with this number increasing after the 2023 cutoff.
This result raises doubts about the current effectiveness of LLMs for the reading list generation task, suggesting that such models rely on memorization and could perform poorly on unseen topics compared to academic search engines.

\section{Conclusion}
We introduced \dataset{}, the largest expert-annotated benchmark dataset for reading list generation.
We conducted experiments with commonly used academic search engines and commercial LLMs.
Empirical results highlight the limited performance of search engines on the task and raise concerns about data contamination and hallucinations in LLMs, notably GPT-4o.
%
%
This work lays the groundwork for developing new methods for reading list generation.
Future directions include exploring Retrieval Augmented Generation (RAG) to address the identified issues in LLMs, and evaluation metrics that consider the ordering of reading lists.


\section{Limitations}
\noindent \textbf{Reading lists content and subjectivity.}
\dataset{} contains reading lists collected from tutorial descriptions presented at ACL main conferences. While we ensured that the lists met our requirements, their content reflects the individual interpretations of the authors. We were unable to manually verify what defines a good reading list for each specific field.
The initial assumption seemed to be that reading all articles in a curated reading list is necessary to fully understand a field. While experts design these lists under the premise that the included articles are valuable for understanding their tutorial sessions, it is possible that only a subset of the references is truly essential. This raises questions about whether all references presented in reading lists are necessary for conveying domain knowledge.
\\

\noindent \textbf{Dataset limitations.}
Although \dataset{} represents the largest collection of expert-curated reading lists, its size is insufficient for training supervised models and is better suited for use as an evaluation set. Future updates will try to address this limitation by incorporating new publications from the ACL Anthology or other potential data sources. However, as tutorial descriptions currently serve as the primary source of reading lists, the dataset's growth remains dependent on the availability of such publications until alternative sources are identified.\\

\noindent \textbf{Queries discrepancies.}
Inter-annotator agreement highlights notable discrepancies between A2 and the other annotators. While these variations were anticipated due to the open-ended nature of our annotation campaign, it would be useful to explore whether they arise from individual annotator biases or differences in expertise.
Understanding this could help the development of systems tailored to different levels of expertise.
\\

\noindent \textbf{Evaluation of article substitutes.}
Lack of evaluation whether the predicted articles can serve as effective substitutes for items in the expert-curated reading lists may hinder models from accurately generating reading lists that align with the standards and comprehensiveness of those created by human experts. Addressing this issue in future research should explore the interchangeability of articles, as multiple sources may cover similar key aspects of a field.\\

\section{Ethical considerations}

\noindent \textbf{Reading lists automatic generation.}
We aim to tackle the automatic generation of reading lists and have conducted experiments using LLMs to generate them. Given the nature of large language models, automated systems based on these technologies could produce lists that misrepresent the impact of an article within a field or even provide incorrect information. To mitigate these risks, additional verification processes should be implemented if such systems would be deployed to a wider audience.
\noindent Additionally, providing easily digestible reading lists for every field may reduce the curiosity to explore the broader literature or engage with articles from other disciplines, potentially limiting their exposure to diverse perspectives and knowledge.\\

\noindent \textbf{Compilation of reading lists.}
The reading lists in \dataset{} are curated by human experts and published within the ACL Anthology. We did not impose any criteria on the individuals compiling these lists or the specific fields of the referenced works when collecting our dataset. Our only selection criterion was the presence of a document that clearly resembled a reading list. \\

\noindent \textbf{Human annotations.}
The annotation campaign for the request queries was carried out by three annotators chosen based on their expertise in NLP. Two of the annotators are authors of this paper, while the third is a member of the same laboratory. The annotation process required approximately 7 hours per annotator, with an average of 5 minutes spent per sample. \\

\noindent \textbf{Reproducibility.}
We provide detailed descriptions of our methodology and make code and data publicly available under MIT license.

\section*{Acknowledgment}
We sincerely thank the reviewers for their valuable insights, constructive feedback, and thoughtful suggestions on this work.
This work is supported by the AID-CNRS NaviTerm project (convention 2022 65 0079 CNRS Occitanie Ouest).

\bibliography{references.bib}

\appendix

\begin{figure*}[ht!]
    \centering
    \includegraphics[width=\linewidth]{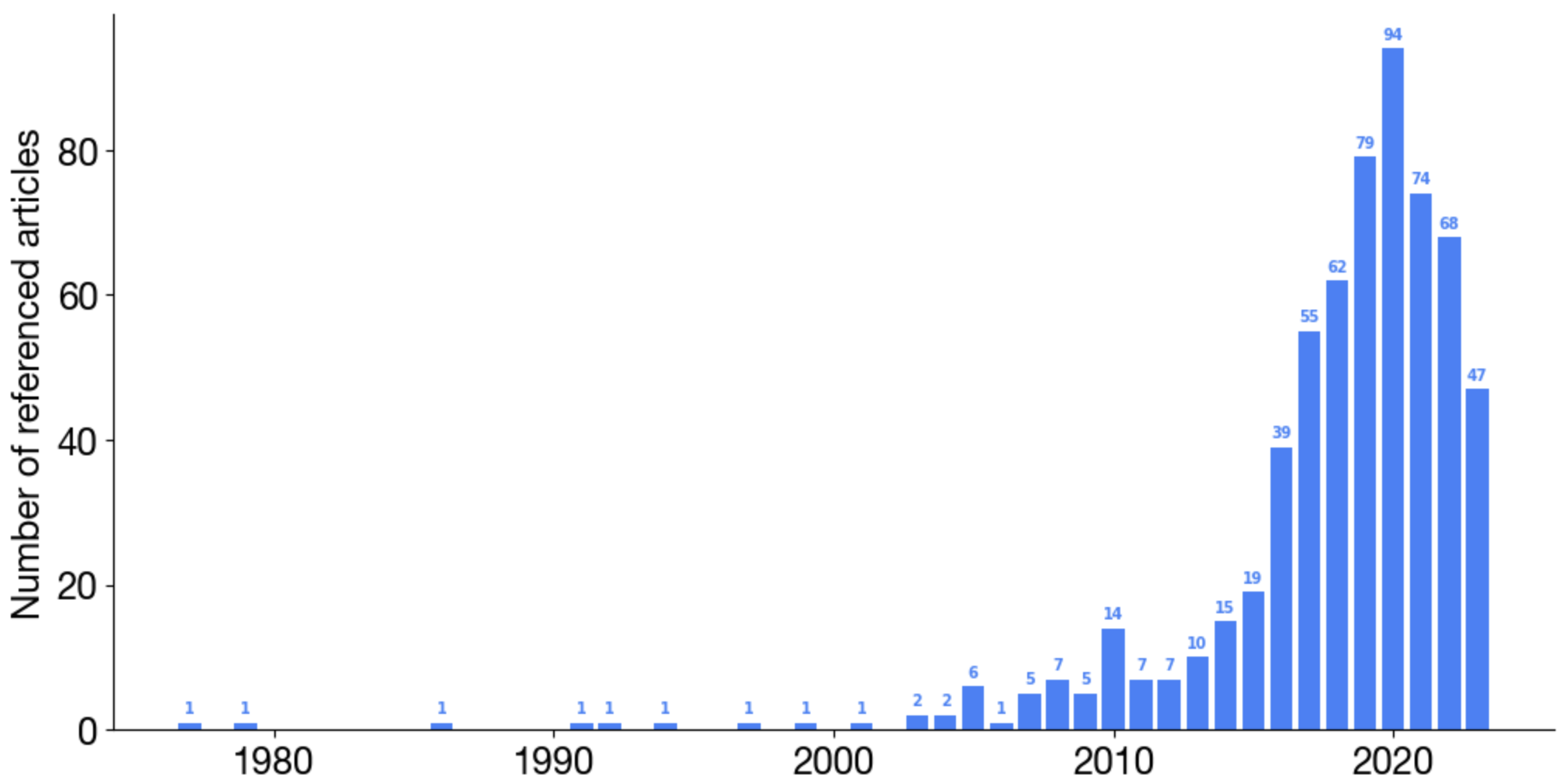}
    \caption{Distribution of the number of referenced articles per year.}
    \label{fig:histogram_year}
\end{figure*}

\begin{table}[htbp]
    \centering
    \input{tables/tutorial_number}
    \caption{Reading lists count with respect to sections sub-levels.}
    \label{tab:section_count}
\end{table}

\begin{table}[htbp]
    \centering
    \input{tables/exp1-sentences}
    \caption{Inter-annotator agreement of natural language instructions.}
    \label{tab:exp1-sentences}
\end{table}

\begin{table*}[htbp]
    \centering
    \resizebox{\linewidth}{!}{%
        \input{tables/current-methods}
    }
    \caption{Performance of Search Engines and LLMs on the reading list generation task.}
    \label{tab:current-methods}
\end{table*}

\begin{table*}[htbp]
    \centering
    \resizebox{\linewidth}{!}{%
        \input{tables/classic-methods}
    }
    \caption{Performance of Retrieval Models on the reading list generation task.}
    \label{tab:classic-methods}
\end{table*}

\section{Annotation Guidelines for creating queries}
\label{appendix:annotation}

\subsection{Objective}

We are developing a dataset aimed at benchmarking the task of automatic reading list generation. Reading lists are curated collections of academic references compiled by experts to assist novice researchers in becoming familiar with a specific research topic.

\subsection{Dataset Source}

Our dataset consists of reading lists extracted from tutorial descriptions presented at ACL conferences.

\subsection{Annotation Task}

Given the title and abstract of a tutorial description on a research topic, your task is to create a search query that you would use to find relevant papers to help you understand that topic.

\subsubsection{Guidelines for creating keyword queries}

    \begin{itemize}
    \item Query structure: each query should be between 1 and 8 keywords long. Keywords should be separated by commas.
    \item Keyword selection: keywords can be words or phrases, keywords are not required to appear in the title and abstract
    \item Relevance: ensure the keywords you choose are highly relevant to the research topic of the tutorial description. Consider what keywords you would use if you were searching for papers to gain an understanding of the topic.
    \item Diversity: use a variety of keywords to capture different aspects of the research topic. Avoid using overly generic keywords that might return irrelevant results.
    \item Specificity: be specific enough to target papers that are closely related to the topic, but not so specific that you exclude important related work.
    When phrase acronyms are highly relevant to a field, please write down the full keyphrase followed by the acronym surrounded by parenthesis. ex: large language models (LLMs)
    \end{itemize}

\subsubsection{Guidelines for creating sentence queries}
    In addition to a keyword query, please annotate natural language sentences that could be used to request an LLM such as ChatGPT about the same information as the keyphrase queries
    \begin{itemize}
    \item Query structure: each query should be between 8 to 30 words and should start with "Give me a reading list on/about"
    \end{itemize}

\section{Protocols for reading lists generation}
\label{appendix:requests}

Due to varying limitations and subtleties among used methods, our procedures differ for each evaluated system. In all cases, we constrained the time range to include articles published on or before the reading list publication.

\subsection{Semantic Scholar}

We use the top 20 results from the Semantic Scholar API’s paper relevance search\footnote{\href{https://api.semanticscholar.org/api-docs/\#tag/Paper-Data/operation/get\_graph\_paper\_relevance_search}{https://api.semanticscholar.org/api-docs/\#tag/Paper-Data/operation/get\_graph\_paper\_relevance\_search}} as a reading list. The paper relevance search proposed by Semantic Scholar involves querying an Elasticsearch index and then re-ranking the top results using a ranker model. 
The request payload is constrained to results within a time range dependent on the publication of the ground-truth reading lists and is configured to return the corpusId, title, and year of predicted articles.

\subsection{Google Scholar}
Because Google Scholar API is not available, we manually crawl the first 20 results from Google Scholar using keyword queries and a specified time range dependent on the publication of the ground-truth reading lists. 
We save the HTML pages for each result and then use regular expressions to extract the article titles and publication years from predicted articles.
After initial information extraction, we request Semantic Scholar API's paper title search\footnote{\href{https://api.semanticscholar.org/api-docs/\#tag/Paper-Data/operation/get\_graph\_paper\_title\_search}{https://api.semanticscholar.org/api-docs/\#tag/Paper-Data/operation/get\_graph\_paper\_title\_search}} to find the corresponding corpusId of each prediction to ensure the article exists in the database.

\subsection{LLMs}
We request the OpenAI\footnote{\href{https://openai.com/api/}{https://openai.com/api/}} and Gemini\footnote{\href{https://ai.google.dev/api}{https://ai.google.dev/api}} APIs using sentence queries that include a specified time range dependent on the publication of the ground-truth reading lists (e.g., "Provide a reading list of 20 articles up to 2023 about..."). The results are saved in markdown format, and we extract the referenced articles from each list using AnyStyle\footnote{\href{https://github.com/inukshuk/anystyle}{https://github.com/inukshuk/anystyle}}. Regular expressions are then applied to identify the titles and publication years of the articles in the predicted lists.
Because this process could lead to parsing errors, another strategy involves conditioning the models to directly generate JSON files as results using JSON-mode or Structured Outputs. 
After the information extraction step, we request Semantic Scholar API's paper title search to find the corresponding corpusId of each prediction and ensure the article exists in the database.

\subsection{BM25}

We implemented BM25S\footnote{\href{https://pypi.org/project/bm25s/}{https://pypi.org/project/bm25s/}} for this method.
We preprocess the queries and ACL Anthology collection by removing stopwords and applying stemming using the pyStemmer library\footnote{\href{https://pypi.org/project/PyStemmer/}{https://pypi.org/project/PyStemmer/}}. From this processed data, we select the top 20 articles relevant to a requested query as the predicted reading list.

\subsection{SPECTER2}
We use SPECTER2 Base\footnote{\href{https://huggingface.co/allenai/specter2\_base}{https://huggingface.co/allenai/specter2\_base}} and its proximity adapter\footnote{\href{https://huggingface.co/allenai/specter2}{https://huggingface.co/allenai/specter2}}.
We embed both the ACL Anthology collection and queries, then calculate cosine similarity using the Faiss library\footnote{\href{https://github.com/facebookresearch/faiss}{https://github.com/facebookresearch/faiss}}. We select the top 20 articles with the highest similarity to a requested query as the predicted reading list.

\end{document}

%% file: tables/datasets-compact.tex
\begin{tabular}{l l l r r c c} 
\toprule
\textbf{Dataset} & \textbf{Source} & \textbf{Domain} & \textbf{\# Lists} & \textbf{\# Items} & \textbf{Expert} & \textbf{Available} \\ 
\midrule
\citet{10.1145/1864708.1864740} & ACM Computing Surveys & CS & 220 & $\geq$15 & \xmark & \xmark \\
\citet{10.1109/JCDL.2019.00011} & Scopus & CS, Engineering & 1\,648 & Unk. & \xmark & \xmark \\
\citet{9835398} & S2ORC, Google Scholar  & NLP, ML, AI & 9\,321 & $\sim$58 & \xmark & \textcolor{teal}{\cmark} \\
\midrule
\citet{UCAM-CL-TR-848} & ACL Anthology Network & NLP & 8 & $\sim$12 & \textcolor{teal}{\cmark} & \textcolor{teal}{\cmark} \\
\textbf{Our dataset} & ACL Anthology & NLP & 85 & $\sim$8 & \textcolor{teal}{\cmark} & \textcolor{teal}{\cmark} \\
\bottomrule
\end{tabular}%

%% file: tables/query-example.tex
\begin{tabular}{@{}l p{.48\textwidth}@{}}
\toprule
   \textbf{Tutorial title} & Computational Analysis of Political Texts: Bridging Research Efforts Across Communities  \\
   \textbf{Abstract} & In the last twenty years, political scientists started adopting and developing natural language processing (NLP) methods more actively in order to [...] \\
\midrule
  \textbf{Query A1}  & analysis of political texts, computational analysis \\[.3em]
  \textbf{Query A2}  & political text analysis, natural language processing, political science, topic detection, stance detection, political text corpus, election prediction \vspace{.3em} \\[.3em]
  \textbf{Query A3}  & political texts, computational analysis \\[.3em]
\midrule
  \textbf{Instr. A3}  & Give me a reading list about computational analysis of political texts. \\
\bottomrule
\end{tabular}

%% file: tables/exp1-keywords.tex
\scalebox{0.8}{\begin{tabular}{lrrr|r}
\toprule
\textbf{Annotator}  & A1 & A2 & A3 &  \textbf{length }\\
\midrule
    A1 (phd) & - & 0.31 & 0.54 & 2.70  \\
    A2 (postdoc) & 0.31 & - & 0.31 & 3.83 \\
    A3 (senior) & 0.54 & 0.31 & - & 2.34 \\
\midrule
    FirstPhrases (auto.) & 0.40 & 0.27 & 0.40 & 2.35   \\
\bottomrule
\end{tabular}}

%% file: tables/current-methods-short.tex
\begin{tabular}{lrrr}
\toprule
\textbf{Systems} & \textbf{Recall@20} & \textbf{NDCG@20} & \textbf{MRR@20} \\
\midrule
S2  & 5.2 & 3.9 & 7.3 \\
GS  & 7.5 & 6.4 & 11.8 \\
\midrule
\textbf{GPT-4o} & \textbf{11.0} & \textbf{10.9} & \textbf{20.5} \\
GPT-4o (Json) & 6.7 &  7.8 & 16.7 \\
GPT-4o (SO) & 6.6 & 7.9 & 16.6 \\
Gemini 1.5 Flash & 3.4 & 3.0 & 6.0 \\
Gemini 1.5 Flash (JM) & 5.9 & 8.1 & 12.3 \\
Gemini 1.5 pro (JM) & 8.2 & 6.9 & 14.1 \\
\bottomrule
\end{tabular}

%% file: tables/classic-methods-short.tex
\begin{tabular}{lrrr}
\toprule
\textbf{Systems} & \textbf{Recall@20} & \textbf{NDCG@20} & \textbf{MRR@20} \\
\midrule
S2 (Any) & 4.5 & 2.5 & 3.3 \\
S2 (ACL) & 7.9 & 5.6 & 8.0 \\
\textbf{S2 (ACL - Most Cited)} & \textbf{10.4} & 
\textbf{7.1} & \textbf{10.3} \\
\midrule
BM25 & 9.7 & 5.9 & 8.1 \\
SPECTER2 & 6.6 & 3.7 & 4.6 \\
\bottomrule
\end{tabular}

%% file: tables/tutorial_number.tex
\begin{tabular}{rrr|r}
\toprule
\multicolumn{3}{c|}{\textbf{\# Lists}} & \textbf{Total} \\
flat  & 1-level & 2-level &  \\
\midrule
 58 & 24 & 3 & 85   \\
\bottomrule
\end{tabular}

%% file: tables/exp1-sentences.tex
\begin{tabular}{lrrr}
\toprule
\textbf{Annotator}  & A1 & A2 & A3\\
\midrule
    A1 (phd) & - & 0.41 & 0.60  \\
    A2 (postdoc) & 0.41 & - & 0.41\\
    A3 (senior) & 0.60 & 0.41 & - \\
\bottomrule
\end{tabular}

%% file: tables/current-methods.tex
\begin{tabular}{lrrrr|rrrr|rrrr}
\toprule
 & \multicolumn{4}{c}{\textbf{Recall@20}} & \multicolumn{4}{c}{\textbf{NDCG@20}} & \multicolumn{4}{c}{\textbf{MRR@20}} \\
 \cmidrule(lr){2-5} \cmidrule(lr){6-9} \cmidrule(lr){10-13} 
 & A1 & A2 & A3 & \textbf{Avg} & A1 & A2 & A3 & \textbf{Avg} & A1 & A2 & A3 & \textbf{Avg} \\
\midrule
S2 & 5.3 & 3.9 & 6.6 & 5.2 & 3.8 & 3.1 & 4.8 & 3.9 & 7.0 & 5.8 & 9.1 & 7.3 \\
GS & 8.4 & 5.2 & 8.8 & 7.5 & 7.2 & 4.7 & 7.1 & 6.4 & 13.0 & 9.7 & 12.7 & 11.8 \\
\midrule
GPT-4o & 11.1 & 10.3 & 11.6 & 11.0 & 11.1 & 9.8 & 11.9 & 10.9 & 20.3 & 17.7 & 23.6 & 20.5 \\
GPT-4o (Json) & 6.9 & 6.4 & 6.8 & 6.7 & 8.4 & 6.8 & 8.2 & 7.8 & 16.4 & 15.0 & 18.7 & 16.7 \\
GPT-4o (SO) & 6.8 & 6.3 & 6.6 & 6.6 & 8.7 & 6.7 & 8.4 & 7.9 & 18.0 & 14.6 & 17.2 & 16.6 \\
Gemini 1.5 Flash & 2.3 & 3.6 & 4.2 & 3.4 & 2.2 & 3.5 & 3.3 & 3.0 & 4.4 & 8.2 & 5.5 & 6.0 \\
Gemini 1.5 Flash (JM) & 6.5 & 5.0 & 6.2 & 5.9 & 9.0 & 6.7 & 8.5 & 8.1 & 13.3 & 12.1 & 11.4 & 12.3 \\
Gemini 1.5 pro (JM) & 7.9 & 7.6 & 9.0 & 8.2 & 7.2 & 5.9 & 7.7 & 6.9 & 14.5 & 11.3 & 16.5 & 14.1 \\
\bottomrule
\end{tabular}

%% file: tables/classic-methods.tex
\begin{tabular}{lrrrr|rrrr|rrrr}
\toprule
 & \multicolumn{4}{c}{\textbf{Recall@20}} & \multicolumn{4}{c}{\textbf{NDCG@20}} & \multicolumn{4}{c}{\textbf{MRR@20}} \\
 \cmidrule(lr){2-5} \cmidrule(lr){6-9} \cmidrule(lr){10-13} 
 & A1 & A2 & A3 & \textbf{Avg} & A1 & A2 & A3 & \textbf{Avg} & A1 & A2 & A3 & \textbf{Avg} \\
\midrule
S2 (Any) & 4.2 & 2.7 & 6.7 & 4.5 & 2.1 & 1.5 & 3.9 & 2.5 & 2.5 & 2.0 & 5.6 & 3.3 \\
S2 (ACL) & 8.4 & 5.2 & 10.2 & 7.9 & 5.5 & 3.7 & 7.5 & 5.6 & 7.6 & 5.3 & 11.2 & 8.0 \\
S2 (ACL - Most Cited) & 10.6 & 5.9 & 14.6 & 10.4 & 7.3 & 4.3 & 9.9 & 7.1 & 11.0 & 6.5 & 13.3 & 10.3 \\
\midrule
BM25 & 8.9 & 9.2 & 11.2 & 9.7 & 5.7 & 5.5 & 6.6 & 5.9 & 8.8 & 6.3 & 9.2 & 8.1 \\
SPECTER2 & 5.8 & 5.7 & 8.2 & 6.6 & 3.1 & 3.4 & 4.7 & 3.7 & 3.3 & 4.6 & 5.8 & 4.6 \\
\bottomrule
\end{tabular}

%% file: main.bbl
\begin{thebibliography}{18}
\providecommand{\natexlab}[1]{#1}

\bibitem[{Boudin(2016)}]{boudin:2016:COLINGDEMO}
Florian Boudin. 2016.
\newblock \href {http://aclweb.org/anthology/C16-2015} {$\texttt{pke:}$ an open
  source python-based keyphrase extraction toolkit}.
\newblock In \emph{Proceedings of COLING 2016, the 26th International
  Conference on Computational Linguistics: System Demonstrations}, pages
  69--73, Osaka, Japan.

\bibitem[{Ding et~al.(2022)Ding, Xiang, Ou, Zuo, Zhao, Lin, Zheng, and
  Liu}]{9835398}
Jiayuan Ding, Tong Xiang, Zijing Ou, Wangyang Zuo, Ruihui Zhao, Chenhua Lin,
  Yefeng Zheng, and Bang Liu. 2022.
\newblock \href {https://doi.org/10.1109/ICDE53745.2022.00322} {{Tell Me How to
  Survey: Literature Review Made Simple with Automatic Reading Path
  Generation}}.
\newblock In \emph{2022 IEEE 38th International Conference on Data Engineering
  (ICDE)}, pages 3426--3438.

\bibitem[{Ekstrand et~al.(2010)Ekstrand, Kannan, Stemper, Butler, Konstan, and
  Riedl}]{10.1145/1864708.1864740}
Michael~D. Ekstrand, Praveen Kannan, James~A. Stemper, John~T. Butler,
  Joseph~A. Konstan, and John~T. Riedl. 2010.
\newblock \href {https://doi.org/10.1145/1864708.1864740} {Automatically
  building research reading lists}.
\newblock In \emph{Proceedings of the Fourth ACM Conference on Recommender
  Systems}, RecSys '10, page 159–166, New York, NY, USA. Association for
  Computing Machinery.

\bibitem[{Figueira et~al.(2019)Figueira, Bel\'{e}m, Almeida, and
  Gon\c{c}alves}]{10.1109/JCDL.2019.00011}
Pablo Figueira, Fabiano Bel\'{e}m, Jussara~M. Almeida, and Marcos~A.
  Gon\c{c}alves. 2019.
\newblock \href {https://doi.org/10.1109/JCDL.2019.00011} {{Automatic
  Generation of Initial Reading Lists: Requirements and Solutions}}.
\newblock In \emph{Proceedings of the 18th Joint Conference on Digital
  Libraries}, JCDL '19, page 1–10. IEEE Press.

\bibitem[{{Gemini Team}(2024)}]{geminiteam2024gemini15unlockingmultimodal}
{Gemini Team}. 2024.
\newblock \href {https://arxiv.org/abs/2403.05530} {{Gemini 1.5: Unlocking
  multimodal understanding across millions of tokens of context}}.
\newblock \emph{Preprint}, arXiv:2403.05530.

\bibitem[{Glava{\v{s}} et~al.(2019)Glava{\v{s}}, Nanni, and
  Ponzetto}]{glavas-etal-2019-computational}
Goran Glava{\v{s}}, Federico Nanni, and Simone~Paolo Ponzetto. 2019.
\newblock \href {https://doi.org/10.18653/v1/P19-4004} {{Computational Analysis
  of Political Texts: Bridging Research Efforts Across Communities}}.
\newblock In \emph{Proceedings of the 57th Annual Meeting of the Association
  for Computational Linguistics: Tutorial Abstracts}, pages 18--23, Florence,
  Italy. Association for Computational Linguistics.

\bibitem[{Gordon et~al.(2017)Gordon, Aguilar, Sheng, and
  Burns}]{gordon-etal-2017-structured}
Jonathan Gordon, Stephen Aguilar, Emily Sheng, and Gully Burns. 2017.
\newblock \href {https://doi.org/10.18653/v1/W17-5029} {{Structured Generation
  of Technical Reading Lists}}.
\newblock In \emph{Proceedings of the 12th Workshop on Innovative Use of {NLP}
  for Building Educational Applications}, pages 261--270, Copenhagen, Denmark.
  Association for Computational Linguistics.

\bibitem[{Jardine(2014)}]{UCAM-CL-TR-848}
James~G. Jardine. 2014.
\newblock \href {https://doi.org/10.48456/tr-848} {{Automatically generating
  reading lists}}.
\newblock Technical Report UCAM-CL-TR-848, University of Cambridge, Computer
  Laboratory.

\bibitem[{Li et~al.(2017)Li, Schijvenaars, and {de Rijke}}]{LI2017666}
Xinyi Li, Bob~J.A. Schijvenaars, and Maarten {de Rijke}. 2017.
\newblock \href {https://doi.org/10.1016/j.ipm.2017.01.005} {{Investigating
  queries and search failures in academic search}}.
\newblock \emph{Information Processing \& Management}, 53(3):666--683.

\bibitem[{Lo et~al.(2020)Lo, Wang, Neumann, Kinney, and
  Weld}]{lo-wang-2020-s2orc}
Kyle Lo, Lucy~Lu Wang, Mark Neumann, Rodney Kinney, and Daniel Weld. 2020.
\newblock \href {https://doi.org/10.18653/v1/2020.acl-main.447} {{S}2{ORC}: The
  {S}emantic {S}cholar {O}pen {R}esearch {C}orpus}.
\newblock In \emph{Proceedings of the 58th Annual Meeting of the Association
  for Computational Linguistics}, pages 4969--4983, Online. Association for
  Computational Linguistics.

\bibitem[{OpenAI(2024)}]{openai2024gpt4technicalreport}
OpenAI. 2024.
\newblock \href {https://arxiv.org/abs/2303.08774} {{GPT-4 Technical Report}}.
\newblock \emph{Preprint}, arXiv:2303.08774.

\bibitem[{Radev et~al.(2013)Radev, Muthukrishnan, Qazvinian, and
  Abu-Jbara}]{radev_acl_2013}
Dragomir~R. Radev, Pradeep Muthukrishnan, Vahed Qazvinian, and Amjad Abu-Jbara.
  2013.
\newblock \href {https://doi.org/10.1007/s10579-012-9211-2} {The {ACL}
  anthology network corpus}.
\newblock \emph{Language Resources and Evaluation}, 47(4):919--944.

\bibitem[{Robertson and Zaragoza(2009)}]{10.1561/1500000019}
Stephen Robertson and Hugo Zaragoza. 2009.
\newblock \href {https://doi.org/10.1561/1500000019} {{The Probabilistic
  Relevance Framework: BM25 and Beyond}}.
\newblock \emph{Found. Trends Inf. Retr.}, 3(4):333–389.

\bibitem[{Sainz et~al.(2023)Sainz, Campos, Garc{\'\i}a-Ferrero, Etxaniz,
  de~Lacalle, and Agirre}]{sainz-etal-2023-nlp}
Oscar Sainz, Jon Campos, Iker Garc{\'\i}a-Ferrero, Julen Etxaniz, Oier~Lopez
  de~Lacalle, and Eneko Agirre. 2023.
\newblock \href {https://doi.org/10.18653/v1/2023.findings-emnlp.722} {{{NLP}
  Evaluation in trouble: On the Need to Measure {LLM} Data Contamination for
  each Benchmark}}.
\newblock In \emph{Findings of the Association for Computational Linguistics:
  EMNLP 2023}, pages 10776--10787, Singapore. Association for Computational
  Linguistics.

\bibitem[{Sesagiri~Raamkumar et~al.(2017)Sesagiri~Raamkumar, Foo, and
  Pang}]{sesagiri_raamkumar_using_2017}
Aravind Sesagiri~Raamkumar, Schubert Foo, and Natalie Pang. 2017.
\newblock \href {https://doi.org/10.1016/j.ipm.2016.12.006} {Using
  author-specified keywords in building an initial reading list of research
  papers in scientific paper retrieval and recommender systems}.
\newblock \emph{Information Processing \& Management}, 53(3):577--594.
\newblock Number: 3.

\bibitem[{Siddall and Rose(2014)}]{Siddall_2014}
Gillian Siddall and Hannah Rose. 2014.
\newblock \href {https://doi.org/10.29173/lirg605} {{Reading lists – time for
  a reality check? An investigation into the use of reading lists as a
  pedagogical tool to support the development of information skills amongst
  Foundation Degree students}}.
\newblock \emph{Library and Information Research}, 38(118):52–73.

\bibitem[{Singh et~al.(2023)Singh, D{'}Arcy, Cohan, Downey, and
  Feldman}]{singh-etal-2023-scirepeval}
Amanpreet Singh, Mike D{'}Arcy, Arman Cohan, Doug Downey, and Sergey Feldman.
  2023.
\newblock \href {https://doi.org/10.18653/v1/2023.emnlp-main.338}
  {{{S}ci{R}ep{E}val: A Multi-Format Benchmark for Scientific Document
  Representations}}.
\newblock In \emph{Proceedings of the 2023 Conference on Empirical Methods in
  Natural Language Processing}, pages 5548--5566, Singapore. Association for
  Computational Linguistics.

\bibitem[{Thompson et~al.(2004)Thompson, Mahon, and Thomas}]{2436/3693}
Lisa Thompson, Claire Mahon, and Linda Thomas. 2004.
\newblock \href {http://hdl.handle.net/2436/3693} {Reading lists - how do you
  eat yours?}
\newblock In \emph{CELT Learning and Teaching Projects 2003/04}. University of
  Wolverhampton.

\end{thebibliography}
